# A fundamental scale of mass for black holes from the cosmological constant

Scott Funkhouser
Dept. of Physics, The Citadel, Charleston, SC 29409

ABSTRACT
The existence of a positive cosmological constant leads naturally to two fundamental scales of length, being the De Sitter horizon and the radius of the cell associated with a holographic degree of freedom. Associated with each of those scales of length are a macroscopic gravitational mass and a microscopic quantum mechanical mass. Three of those four fundamental masses have been discussed in the literature, and this present work identifies the physical significance of the remaining mass, being the gravitational mass associated with the holographic length. That mass, which is of the order $10^{12}$kg and inversely proportional to the sixth root of the cosmological constant, represents the mass of the black hole whose evaporation time is equal to the fundamental cosmic time, which is of the order the current age of the universe. It also represents the minimum mass of a black hole that is capable of accreting a particle whose Compton wavelength is equal to the fundamental holographic length, which is of the order the Compton wavelength of the nucleon.

*1. Introduction*

In this cosmological epoch the inventory of energy in the observable universe has begun apparently to be dominated by some form of vacuum energy. The presence of uniform vacuum energy, whose density is empirically near $6 \times 10^{-10}$J/kg$^3$, is consistent with the existence of a cosmological constant, $\Lambda$, that is approximately $1 \times 10^{-35}$s$^{-2}$. In future ages, as the density of matter continues to decrease, the universe will approach asymptotically a De Sitter space in which the only form of energy is the cosmological vacuum energy. The largest possible event horizon in a universe with a positive cosmological constant is the De Sitter horizon $R_\Lambda$, given by

$$R_\Lambda = c\left(\frac{3}{\Lambda}\right)^{1/2}, \qquad (1)$$

where $c$ is the vacuum-speed of light. The fundamental, cosmological time $t_\Lambda$ that follows from the cosmological constant may be defined as

$$t_\Lambda \equiv \frac{R_\Lambda}{c} = \left(\frac{3}{\Lambda}\right)^{1/2}, \qquad (2)$$

which is of the order $10^{17}$s. That time is roughly the age of the universe in which the energy density that results from the cosmological constant is of the order the energy density of the cosmic matter. It is regarded as a remarkable coincidence that the current age of the universe is of the order $t_\Lambda$. For proper times much larger than $t_\Lambda$ the mass contained within the sphere whose radius is the event horizon dwindles, eventually rendering observational cosmology ineffective. [1], [2]

The existence of a positive cosmological constant leads naturally to at least four fundamental, physically significant scales of mass. Each of the masses addressed in this present investigation may be defined naturally in terms of a characteristic scale of length. The largest charge-neutral, non-rotating mass that could be contained within a sphere whose radius is $R$ is the mass of a Schwarzschild black hole (SBH) whose horizon is $R$, and is given by

$$m_G(R) \equiv \frac{c^2}{2G}R, \qquad (3)$$

where $G$ is the Newtonian gravitational coupling. The smallest possible mass that could be contained within a sphere of radius $R$ is approximately the mass of the particle whose Compton wavelength is equal to $2R$, and is given by

$$m_h(R) \equiv \frac{h}{2Rc} = \frac{\pi\hbar}{Rc}, \tag{4}$$

where $h \equiv 2\pi\hbar$ is the Planck quantum. The term $m_G(R)$ is greater than $m_h(R)$ for all radii $R$ larger than $(2\pi)^{1/2}l_P$, where $l_P$ is the Planck length. Note that geometrical and other unitless numerical coefficients of order near 10 are not important in this present analysis.

The De Sitter horizon represents the fundamental cosmic scale of distance associated with the cosmological constant, and it is therefore appropriate to examine the two masses that follow from (3) and (4) for $R=R_\Lambda$. The largest mass that could be contained within a sphere whose radius is $R_\Lambda$ is given by

$$M_\Lambda \equiv m_G(R_\Lambda) \sim \frac{c^3}{G\Lambda^{1/2}}. \tag{5}$$

The mass $M_\Lambda$ is physically significant in the standard model of cosmology. The mass $M_e(t)$ contained within a sphere whose radius is the event horizon approaches $AM_\Lambda$ in the limit of vanishing proper time $t$, where $A$ represents a numerical factor. The mass $M_p(t)$ contained within the sphere whose radius is the particle horizon approaches the same upper bound $AM_\Lambda$ as $t$ increases. Furthermore, the masses $M_e(t)$ and $M_p(t)$ are both of order near $M_\Lambda$ for $t \sim t_\Lambda$. The minimum mass that could be contained within a sphere whose radius is $R_\Lambda$ is of the order

$$w_\Lambda \equiv m_h(R_\Lambda) \sim \frac{\hbar\Lambda^{1/2}}{c^2}. \tag{6}$$

The wavelength of any quantum of energy smaller than $w_\Lambda c^2$ would be larger than the De Sitter horizon, and $w_\Lambda c^2$ represents therefore the minimum possible quantum of energy in a universe with a positive cosmological constant. The mass in (6) is known as the Wesson mass, which may represent the fundamental scale by which all matter is quantized [3].

There exists another scale of length that is associated with the cosmological constant and from which two additional, physically significant scales of mass may be obtained. It follows from holographic principles and the thermodynamics of black holes that the maximum number of bits $N(R)$ that could be registered by the contents of a sphere whose radius is $R$ is [4]

$$N(R) = \frac{\pi R^2}{\ln(2)l_P^2}. \tag{7}$$

The thermodynamic entropy of a body is proportional to the number of bits registered by the body. The number of bits registered by a SBH whose horizon is $R_1$ is equal to the maximum number allowed to a sphere whose radius is equal to $R_1$ [4]. The entropy of the black hole represents therefore the maximum possible entropy that could be associated with the contents of a sphere of radius $R_1$.

The maximum number $N_\Lambda \equiv N(R_\Lambda)$ of bits that could be registered by a sphere whose radius is the De Sitter horizon is

$$N_\Lambda = \frac{3\pi}{\ln(2)} \frac{c^5}{G\hbar\Lambda}. \tag{8}$$

The pure number in (8), which is of the order $10^{122}$, represents the maximum number of bits that could be registered by a universe with a positive cosmological constant [1]. The volume of a De Sitter space divided by the maximum number $N_\Lambda$ of registered bits defines a characteristic, microscopic volume $v_\Lambda$ given by

$$v_\Lambda = \frac{4\pi R_\Lambda^3/3}{N_\Lambda} = \frac{4\ln(2)}{3} R_\Lambda l_P^2. \qquad (9)$$

The radius $r_\Lambda$ of the sphere whose volume is $v_\Lambda$ is given by

$$r_\Lambda \sim \left(\frac{G\hbar}{c^2}\right)^{1/3} \Lambda^{-1/6}. \qquad (10)$$

The fundamental length $r_\Lambda$ is of the order $10^{-15}$m, and it is expected to establish a fundamental, physically significant scale in our universe [5], [6]. The smallest possible mass that could be contained within a sphere whose radius is $r_\Lambda$ is given by

$$m_\Lambda \equiv m_h(r_\Lambda) \sim \left(\frac{\hbar^2}{Gc}\right)^{1/3} \Lambda^{1/6}, \qquad (11)$$

which is also approximately the mass of the particle whose Compton wavelength is $r_\Lambda$. The fundamental holographic mass $m_\Lambda$ is of the order $10^{-28}$kg (since $r_\Lambda$ is of the order the Compton wavelength of the nucleon). A variety of independent but consistent theories indicate that the remarkable similarity between $m_n$ and $m_\Lambda$ is the result of a physical relationship [1],[5],[6],[7],[8].

The largest possible mass that could be contained within a sphere whose radius is the holographic length $r_\Lambda$ is given by

$$\mu_\Lambda \equiv m_G(r_\Lambda) \sim \left(\frac{\hbar c^4}{G^2}\right)^{1/3} \Lambda^{-1/6}, \qquad (12)$$

which is of the order $10^{12}$kg. The fundamental mass $\mu_\Lambda$ is the primary subject of this present investigation, and in the following section its physical significance is established.

*2. A fundamental scale for black holes*

According to Hawking's analysis of the behavior of quantum fields in vicinity of black holes, the surface just above the horizon of a black hole radiates thermally. The spectrum of Hawking radiation generated by a SBH is identical to the spectrum of an ideal blackbody whose characteristic temperature $T(M)$ is determined by the mass $M$ of the black hole according to

$$T(M) = \frac{1}{k_B} \frac{\hbar c^3}{8\pi GM}, \qquad (13)$$

where $k_B$ is the Boltzmann constant [9], [10]. The Hawking radiation should cause effectively the black hole to lose energy and thus also mass. Unless a given black hole should continuously accrete energy, it must evaporate eventually due to its inherent luminosity. If a SBH of mass $M$ does not accrete significant quantities of energy it will evaporate after a time $t(M)$ given by

$$t(M) = \frac{5120\pi G^2 M^3}{\hbar c^4}. \qquad (14)$$

Equation (14) may be stated alternatively as

$$M(t) = \left(\frac{\hbar c^4 t}{5120\pi G^2}\right)^{1/3}, \tag{15}$$

where $M(t)$ is the mass of a SBH whose evaporation time is $t$ [9], [10].

Since the preponderance of cosmic mass $M_\Lambda$ recedes beyond the event horizon for proper times much greater than $t_\Lambda$, the greatest possible time for which a body could be causally connected to the bulk of the observable universe is of order near $t_\Lambda$. Barring significant accretion, the mass of a SBH that could exist for a time $t_\Lambda$ is given by

$$M(t_\Lambda) = \left(\frac{1}{5120\pi}\right)^{1/3}\left(\frac{\hbar c^4}{G^2}\right)^{1/3}\Lambda^{-1/6}. \tag{16}$$

The right side of (16) differs from $\mu_\Lambda$ only by a geometrical coefficient. The fundamental mass $\mu_\Lambda$ represents therefore the minimum mass of a primordial black hole that could be connected causally to the bulk of the observable universe for a time $t_\Lambda$. Note also that the age of the universe is currently of the order $t_\Lambda$. Barring significant accretion, the minimum mass of a primordial black hole that could be observed in this special epoch is therefore of the order $M(t_\Lambda) \sim \mu_\Lambda$.

There is another physical significance associated with $\mu_\Lambda$ that follows from considering the accretion of a particle by a black hole. If the quantum wavelength of the particle should be of the order the horizon of the black hole or larger, then the classical treatment of the accretion process ceases to be valid, and the probability of accretion may be attenuated substantially. It follows from (3) and (4) that the Compton wavelength of a particle whose mass is $m_\Lambda$ is of the order the horizon of a SBH whose mass is $\mu_\Lambda$. The fundamental mass $\mu_\Lambda$ represents therefore the minimum mass of a black hole that could classically accrete a particle of mass $m_\Lambda$. Since $m_\Lambda$ is of order near $m_n$, perhaps as a result of a physical relationship, $\mu_\Lambda$ is approximately the minimum mass of a SBH that could accrete a nucleon. [11]

It is instructive to consider two additional relationships that follow from the present considerations. If a SBH whose initial mass is $M$ absorbs a particle of mass $m$, where $m \ll M$, then the temperature of the black hole changes by an amount $\Delta T(M,m) = T(M+m) - T(M)$, which is approximately $-T(M)m/M$. (Let $\Delta T(M,m)$ be defined as 0 for any $M,m$ for which accretion is impossible or strongly attenuated.) The temperature $T(M)$ of a SBH whose mass is $M = \mu_\Lambda$ is given by

$$T(\mu_\Lambda) \sim \frac{1}{k_B}\left(\frac{\hbar^2}{Gc}\right)^{1/3}\Lambda^{1/6}, \tag{17}$$

and thus

$$k_B T(\mu_\Lambda) \sim m_\Lambda c^2, \tag{18}$$

which is of order near the rest energy of the nucleon. The change in the temperature of a SBH of mass $\mu_\Lambda$ upon accreting a particle of mass $m = m_\Lambda$ is

$$\Delta T(\mu_\Lambda, m_\Lambda) \cong -\frac{1}{k_B}\frac{m_\Lambda^2}{\mu_\Lambda}c^2, \tag{19}$$

which reduces to

$$k_B \Delta T(\mu_\Lambda, m_\Lambda) \cong -w_\Lambda c^2, \tag{20}$$

which is the minimum possible quantum of energy in a universe with a positive cosmological constant. Note that the temperature of a SBH that is more massive than $\mu_\Lambda$ changes by an amount whose magnitude is smaller than $w_\Lambda c^2$ upon absorbing a particle whose mass is $\mu_\Lambda$, which is of order near the nucleon mass. Since black holes more massive than $\mu_\Lambda$ presumably accrete nucleons, $k_B w_\Lambda c^2$ must not represent the lower bound on the magnitude of the change in temperature of a black hole upon accreting a nucleon. In fact, $k_B w_\Lambda c^2 \sim \Delta T(\mu_\Lambda, m_\Lambda)$ represents the *upper* bound on the magnitude of $\Delta T(M, m_\Lambda)$ since $\mu_\Lambda$ is the minimum mass of a SBH that could absorb a particle of mass $m_\Lambda$, and the magnitude of $\Delta T(M,m)$ increases as $M$ decreases, for any given $m > h/(Rc)$.